\documentclass[runningheads]{llncs}
\usepackage{graphicx}
\usepackage{multirow}
\usepackage{subfigure}

\begin{document}
\title{Experiences from Using LoRa and IEEE 802.15.4 for IoT-enabled Classrooms\thanks{This work has been supported by the EU research project ``European Extreme Performing Big Data Stacks'' (E2Data), funded by the European Commission under H2020 and contract number 780245, and by the ``Green Awareness In Action'' (GAIA) research project, funded by the European Commission and the EASME under H2020 and contract number 696029. This document reflects only the authors' views and the EC and EASME are not responsible for any use that may be made of the information it contains. \textbf{Preprint version of the paper submitted to 2019 European Conference on Ambient Intelligence, 13-15 November 2019, Rome, Italy. AmI 2019. Lecture Notes in Computer Science, vol 11912. Springer, Cham. https://doi.org/10.1007/978-3-030-34255-5\_13}
}}
\titlerunning{Experiences from Using LoRa \& IEEE 802.15.4 for IoT-enabled Classrooms}
%
\author{Lidia Pocero\inst{1} \and
Stelios Tsampas\inst{1} \and
Georgios Mylonas\inst{1} \and
Dimitrios Amaxilatis\inst{1}}
\authorrunning{ }
%
\institute{Computer Technology Institute and Press ``Diophantus'', Rio, Patras, Greece
\email{\{pocero, tsampas, mylonasg, amaxilat\}@cti.gr}\\
}
\maketitle              
\begin{abstract}
Several networking technologies targeting the IoT application space currently compete within the smart city domain, both in outdoor and indoor deployments. However, up till now, there is no clear winner, and results from real-world deployments have only recently started to surface. In this paper, we present a comparative study of 2 popular IoT networking technologies, LoRa and IEEE 802.15.4, within the context of a research-oriented IoT deployment inside school buildings in Europe, targeting energy efficiency in education. We evaluate the actual performance of these two technologies in real-world settings, presenting a comparative study on the effect of parameters like the built environment, network quality, or data rate. Our results indicate that both technologies have their advantages, and while in certain cases both are perfectly adequate, in our use case LoRa exhibits a more robust behavior. Moreover, LoRa's  characteristics make it a very good choice for indoor IoT deployments such as in educational buildings, and especially in cases where there are low bandwidth requirements.

\end{abstract}
%
%
\section{Introduction}

The smart cities and the Internet of Things (IoT) domains are currently among the most active research areas, having gradually progressed from being mere buzzwords to having actual large-scale installations deployed and applications developed. In this context, a number of competing wireless networking technologies have surfaced in recent years, aiming to appeal to the communities that engage within these two domains. Advancements in wireless communications technology have enabled a multitude of different approaches to the trade-off between power consumption, communication range and bandwidth, in order to answer to all the various types of application use-case requirements. In this context, recent technologies like LoRaWAN and NB-IoT have surfaced, aiming to claim a place in the area originally covered by technologies like ZigBee.

As part of our research activity, we have developed an IoT platform that combines sensing, web-based tools and gamification elements, in order to address the educational community. Within the context of a research project, its aim is to increase awareness about energy consumption and sustainability, based on real-world sensor data produced by the school buildings where students and teachers live and work, while also lead towards behavior change in terms of energy efficiency. This real-world IoT deployment developed through the aforementioned project provides real-time monitoring of 25 school buildings spread in 3 European countries. 

Due to the multi-year development phase of the project, a number of conditions, like limited availability of certain networking components and appearance on the market of new ones, have led us to follow a heterogeneous approach with several networking technologies utilized in different buildings of our deployments. During the previous development phases, we have used almost exclusively IEEE 802.15.4-based 2.4GHz modules. However, school buildings have certain characteristics that in practice lead to less than optimal results in terms of reliability and connectivity. For this reason, we decided to shift towards LoRa-based modules for our deployments in some specific school buildings. LoRa is also well-suited to application use-cases where devices mostly transmit data to the cloud or a nearby gateway (uplink), versus downlink, which also reflects better in the design of other, higher-level, protocols used in IoT. ZigBee and other technologies are better suited for use-cases with more symmetric bandwidth requirements. The frequencies used in LoRa aim for longer range, while they also help to provide a higher degree of wall penetration than other protocols, although 802.15.4 modules are also available in similar frequencies (i.e., apart from 2.4GHz) but their availability is limited and not guaranteed.

In this paper, we present a comparison between LoRa and IEEE 802.15.4 as a networking backbone of an IoT deployment inside a number of school buildings in Europe. We relay our experiences from using both technologies in practice, to develop IoT real-world, reliable and well-performing deployments as a foundation for pervasive computing applications. We present an overview of the two technologies and how we used them in our use-case, along with an analysis of the effect of changing parameters like network density, application data rate and distance between nodes. Our results indicate that in our use-case and under the design constraints that we had, LoRa works in a more reliable manner while also satisfying our data rate requirements.

\section{Previous Work}

Regarding recent comparisons between protocols used for low power wide area networks (LPWAN) in IoT, \cite{comparative-lpwan} and \cite{survey-lpwa} discuss aspects related to LoRa, NB-IoT and ZigBee. All of these technologies are being used especially in smart city applications, and currently there is a lot of interest in understanding the parameters related to the their performance in the real world. This aspect is discussed in \cite{ref1}, where a smart city deployment using LoRa and IEEE 802.15.4 is evaluated using mostly simulation methods and limited real-world studies. \cite{ref8} provides a survey of LoRaWAN for IoT and recent examples of related applications, along with a discussion on its advantages and shortcomings.

Although most related works describing aspects like the ones mentioned above are limited to simulation, there are some recent ones performing measurements in real-world settings. LoRa performance is explored to a certain degree in \cite{ref2}, with a discussion on possibilities and limitations. In \cite{ref3}, a performance evaluation of LoRaWAN and its integration in IoT devices is discussed, while \cite{ref4} explores its scalability in the context of large-Scale sensor networks. \cite{ref5}~presents an evaluation of LoRaWAN using a permanent outdoor deployment, while \cite{ref6} provides an experimental study on LPWANs for mobile IoT applications. \cite{ref7} provided a study of LoRa in long-range use-cases and produced certain radio propagation models to be used when designing LoRa-based solutions. Their work confirmed coverage of up to 8 kilometers in urban and 45 kilometers in urban areas (in line-of-sight conditions). \cite{ref9-SINHA201714} provided a simulation-based comparative study between LoRa and NB-IoT, describing the advantages of each technology in specific areas and use-cases.

However, so far most works are either mostly based on simulation, or they do not attempt a straight apples-to-apples comparison between different networking technologies in specific use-cases e.g., for IoT, pervasive computing or smart cities. Our work here contributes to the discussion over which technology is better suited for real-world application in a representative use-case; school building are a characteristic and ubiquitous example of public building. Our application requirements in terms of data sampling and quality of service (QoS) are also similar to other related application scenarios (e.g., office building monitoring and automation).

\section{Short Overview of IEEE 802.15.4 and LoRa}

In this section, we present a brief comparison between the IEEE 802.15.4 and LoRa  networking, in order to give a context for the sections that follow and discuss their performance in more detail.

\subsection{IEEE 802.15.4}

The IEEE 802.15.4 is a standard for wireless communication. It specifies the use of Direct Sequence Spread Spectrum (DSSS) and an Offset Quadrature Phase Shift Keying (O-QPSK). The IEEE 802.15.4 protocol specification includes both a Physical and a MAC layer definition. The physical layer defined the frequency (possible frequencies are $868MHz$, $915MHz$ and $2.4GHz$) and the number of channels. The MAC layer defines the device types (physical address) and channel access. The 802.15.4 physical layer defines the possibility of $16$ channels in ISM band from $5MHz$ channel spacing, beginning at $2405MHz$ and ending at $2480MHz$. The carrier-sense multiple access with collision avoidance (CSMA/CA) protocol is implemented as part of the MAC layer by using a CCA (clear channel assessment) technique to determine if the channel is available before to transmitting a packet \cite{digi}. 

Moreover, the European Telecommunication Standards Institute (ETSI) regulates the maximum transmitted RF power in wireless networking modules via the ETSI EN~300~328 standard. Two clauses are the most important: the maximum transmit power, which limits power to $100mW$, and the maximum EIRP spectral density, which is limited to $10mW/Hz$~\cite{digi}. The ETSI standard sets a safe limit for RF output power around $12dBm$ \cite{etsi}. Furthermore, in the $2.4GHz$ band, a maximum over-the-air data rate of $250kbps$ is specified, but due to the overhead of the protocol, the actual theoretical maximum data rate is approximately half of that~\cite{seeed}. 

For our network implementation, for the 802.15.4 part we have chosen to use XBee network modules; in the rest of the text, XBee refers to 802.15.4 aspects. We set every XBee module at the 802.15.4 MAC mode with ACKs acknowledgment protocol. The RF module operates in a unicast mode that supports retries. The receiving modules send an ACK of RF packets to confirm reception to the transmitter. If the transmitting module does not receive the ACK, it will resend the packet up to three times, or until the ACK packet is received. The transmission happens directly without any delays. The modules are configured to operate with a peer-to-peer network topology with no master/slave relationship and each module of the network shares both roles master and salve. The Network ID and Channel must be identical across all the modules in the network. Each RF packet contains a maximum of 100 characters ($100bytes$). In our network, the payload of the RF packet will be variable but always smaller than the 100 character limit, which means all messages are transmitted within one packet. 

\subsection{LoRa}

Long Range (LoRa) was originally conceived as a long-range wireless communication technology that operates on the sub-GHz license free ISM bands ($868MHz$ in Europe and $915MHz$ in the U.S.). This means that, in contrast to other related technologies like NB-IoT, it operates in frequencies that are free to use and anybody can potentially operate a LoRa network without requiring a license for it. Regarding features of LoRa that are examined in this work, the over-the-air LoRa modulation technique can be understood as a MFSK modulation on top of a Chirp Spread Spectrum (CSS) method. Each bit is spread by a chipping factor, with the number of chips per bit called Spread Factor (SF). Chirps are used to encode data in LoRa networks for transmission, while inverse chirps are used on the receiver side for signal decoding. The modulation across the channel is weeping so that the transmission signal occupies the chosen bandwidth (BW). SFs specifically set the data transfer rate relative to the range, by essentially indicating how many chirps are used per second, and define bit rates, per symbol radiated power, and achievable range. The possible values of SF are between $6$ and $12$. The data rate depends on the selected SF, e.g., $SF9$ is $4$ times slower than $SF7$ in terms of bit rate. In general, the slower the bit rate, the higher the energy per data set and the higher the range~\cite{ref3}. 

The $868$ ISM frequency band ranges from $865MHz$ to $870MHz$ and is regulated for the European zone\cite{Lavric2017ALL}. The rules are based on two restrictions: a) the maximum power transmission that can be used on a channel at the communication is $25mW$ (equivalent of $14dB$); b) the duty cycle that is defined as the ratio of maximum time-on-air (ToA) per hour and is limited to $1\%$, which in practice restricts the communication of each LoRa device with other nodes to $36$ seconds per hour. The MAC layer of LoRa does not implement any listen-before-talk (LBT) or CSMA to avoid collisions. Instead it implements a pure Aloha protocol, sending data whenever available, thus the number of collisions increases together with transmission rate or network node density.  

\section{A large-scale IoT infrastructure inside school buildings}

\begin{figure}[tb]
\centering
\includegraphics[width=0.98\columnwidth]{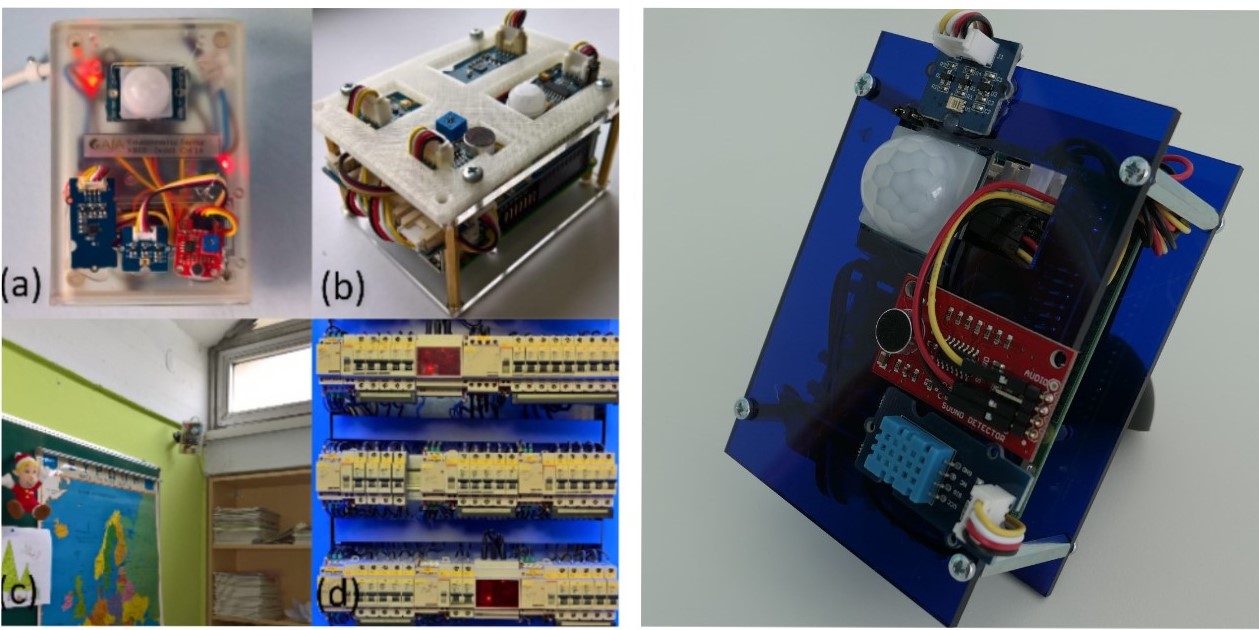}
\caption{Examples of the IoT infrastructure located inside school buildings in Greece (a-b) IoT nodes based on Arduino and Raspberry Pi, c) actual node inside a classroom, d) power meter installed on a electricity distribution panel. on the right part of the figure, the latest hardware revision of the actual environmental nodes used in our IoT infrastructure, utilizing a LoRa communication module.}
\label{fig:nodes}
\end{figure}

Overall, the deployed devices provide $1250$ sensing points organized in four categories: (1) classroom environmental sensors; (2) atmospheric sensors (outdoors); (3) weather stations (on rooftops); and (4) power consumption meters (attached to electricity distribution panels). Given the diverse building characteristics and usage requirements, deployments vary between schools (e.g., number of sensors, manufacturer, networking, etc.). The IoT devices (Fig.~\ref{fig:nodes}) used are either open-design IoT nodes, or off-the-shelf products from IoT device manufacturers. Indoor devices use IEEE 802.15.4 or LoRa wireless networks. These devices are connected to cloud services via IoT gateway devices, which coordinate communication with the rest of the platform, while outdoor nodes use wired networking or WiFi.

\subsection{IEEE 802.15.4 Network Topology}

The IEEE 802.15.4 communication between the IoT nodes is provided by XBee modules connected to each IoT node operated by the Arduino XBee~\cite{ref_url1} and XBeeRadio~\cite{ref_url2} software libraries. Node-to-node communication includes the checksum of the payload, which is validated at the network level for each node to determine erroneous or invalid messages, which are discarded.

\begin{figure}[htbp]
\centerline{
\includegraphics[width=0.48\columnwidth]{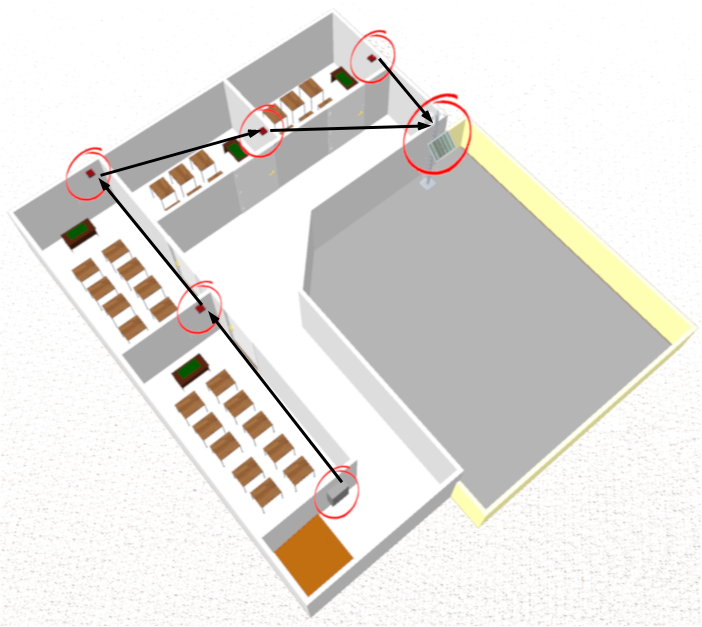}
~
\includegraphics[width=0.48\columnwidth]{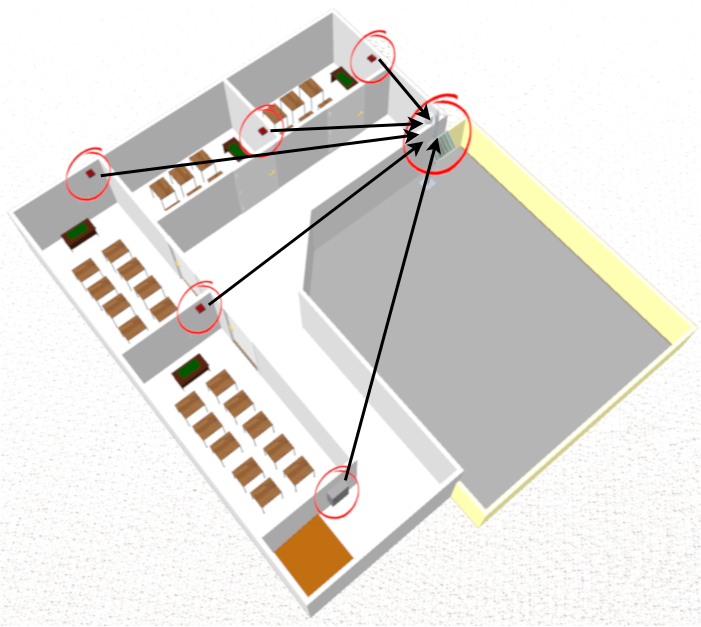}
}
\caption{Examples of data collection routes in an IEEE 802.15.4 (left) and a LoRa (right) IoT deployment, using a tree/multihop and a star network topology, respectively.}
\label{fig:LoRa-xbee-setup}
\end{figure}

All IoT nodes form ad-hoc networks and report their measurements through the designated IoT gateways. Because IEEE 802.15.4 is a short-range communication technology and end-to-end communication is not possible due to power limitations and propagation obstacles, all indoor IoT nodes form an ad-hoc overlaying multi-hop bidirectional tree network. The gateway is the root of this tree and the orchestrator of the network. New nodes can join the network at any time either directly below the gateway, or as a child of the node that is closer to the gateway and has a received signal strength indication (RSSI) lower than a specific threshold ( in our case $90db$). The resulting routing tree allows for bidirectional communication between the IoT nodes and the gateway. The routing library developed for the Arduino and XBee devices is also available on GitHub. An example of a formed network can be seen in Fig.~\ref{fig:LoRa-xbee-setup}(a). Once the network has been established, each node collects environmental or other sensor data and emits a data packer (e.g., an \textit{Environmental Data Packet (EDP)}) to the GW every 10 seconds. The payload size varies depending on the sensing activity, but in our case it is always lower than the limit of a 100 character payload to fit in a single packet. In addition, each environmental node checks its motion (PIR) sensor every 2 seconds and emits a \textit{PIR Data Packet (PDP)} independently each time motion is detected.

\subsection{LoRa Network Topology}
Our IoT nodes based on LoRa use a single-hop topology to cover the necessary distance (tested with up to 3-floor concrete-built buildings), thanks to the communication range and signal penetration characteristics. An example of a formed network showing the difference with the IEEE 802.15.4 network can be seen in Fig.~\ref{fig:LoRa-xbee-setup}(b).
A network, in which the IoT nodes communicate directly with the IoT GW has been installed inside multiple school buildings. We use LoRa to build our own wireless LoRa Personal Area Network (PAN) with a star topology. The IoT nodes communicate using the Grove LoRa $868MHz$ \cite{seeed} modules with a LG01-N Single Channel LoRa Dragino gateway~\cite{dragino}, suitable for small-scale LoRa networks. The communication device for both the Dragino and the Grove modules is based on the RF95 SX1276 LoRa module~\cite{ref_url6}.

The GW coordinates the communication with every node to guarantee that the nodes do not occupy the medium at the same time. This implementation is necessary to avoid the interference due to ALOHA MAC protocol and guarantee no interference between the nodes at the same network. The network is created by the GW announcing itself through broadcast messages. The new nodes reply to the broadcast with a connection request which, if it is accepted by the GW, is acknowledged by a confirmation message. The ALOHA protocol with a randomized delay is used by the nodes to answer to the GW network announcement. 

Once the network is setup, the GW requests data periodically from each node in a sequential fashion with a \textit{Data Request Packet (DRP)}. The nodes reply with a a \textit {Data Packet (DP)} consisting of the sensor measurements. The request rate of the GW is configurable to adjust to the requirements of ToA EU regulations.  In this case, the node samples the PIR sensor between GW requests and includes the motion sensing information in the DP to avoid creating overhead. If a reply is not received or the reply is corrupted, the GW can repeat the DRP up to three times for each node. We implemented our own \textit{CRC (Cycle Redundancy Check)} method at network level to detect message corruption instead of using the LoRa module functionality at MAC level. The network is refreshed every 15 minutes. On each refresh, new nodes can be attached while unreachable nodes are not removed to speedup future reconnects.

\section{LoRa network configurations comparison}
The maximum \textit{ToA} is restricted in Europe, thus limiting the packet rate for each network device. The ToA of each LoRa packet depends on the \textit {spreading factor} (SF), \textit{coding rate} (CR), \textit {signal bandwidth}  (BW) and the \textit{packet payload} (PL). The LoRa packet duration is the sum of the duration of the preamble and the transmitted packet. The data-sheet of  the SX1276 module \cite{ref_url6} describes the formula to calculate the number of payload symbols and the preamble length. We use this to determine the ToA of each packet of our network in milliseconds, and thus we can calculate the maximum legal packet rate to accommodate the $36$ seconds ToA per node limit under different network configurations. The maximum \textit{PL} of each \textit{DP} is $60 bytes$ and the \textit{PL} of each \textit{DRP} is fixed at $4 bytes$. A comparison for each type of packet between different network configuration is described in Table~\ref{tab1} showing the corresponding \textit{PL}, \textit{ToA} for a single packet in milliseconds and the minimum \textit{Period} between transmissions in seconds. %

\begin{table}
\begin{center}
\caption{Packet ToA and Period per device}\label{tab1}
\begin{tabular}{|c|c||c|c|c||c||c|c|}
\hline
\multicolumn{2}{|c||}{} &  \multicolumn{3}{|c||}{Node (Data Packets)} &  \multicolumn{3}{|c|}{Gateway (Data Request Packets)}\\
\hline
SF & BW  & PL & {\bfseries Packet ToA} & {\bfseries Min. Period} & PL & {\bfseries Packet ToA} & {\bfseries Min. Period} \\
 & [kHz] & [Bytes] & [ms]  & [s] & [Bytes] & [ms] & [s]\\
\hline
7 &  125 & 60 & 112.896 & 11.289 & 4& 30.976 & 3.097\\
7 &  250 & 60 & 56.448  & 5.644 & 4 & 15.488 & 1.548\\
7 &  500 & 60 & 25.088  & 2.8224 & 4 & 7.744 & 0.774\\
9 &  125 & 60 & 319.488  & 36.966 & 4 & 123.904 & 12.390\\
9 &  250 & 60 & 159.744  &18.483 & 4 & 61.952 & 6.195\\
9 &  500 & 60 & 79.872  & 9.241 & 4 & 30.976 & 3.097\\  
\hline
\end{tabular}
\end{center}
\end{table}

As described, the GW requests data from each node periodically by sending a \textit{DRP} with a $4 bytes$ payload. As a consequence, in our case the maximum packet rate (minimum period between \textit{DPs}) per node is limited by the maximum packet rate (minimum period between \textit{DRPs}) of the GW, which depends on the total number of nodes in the network. Table~\ref{tab2} presents the theoretical minimum \textit{period} per node and the maximum \textit{packets} per 15 minutes in our network as influenced by the restrictions of the \textit{ToA} of the GW under different network configurations for two schools (\textit{LoRa School A}, \textit{LoRa School B}) of our installation. 

\begin{table}
\begin{center}
\caption{Theoretical minimum Period and maximum Packets per 15 min. \textbf{\textit{n}} represents the number of nodes in the network}\label{tab2}
\begin{tabular}{|c|c||c|c||c|c||c|c|}
\hline
& & \multicolumn{2}{|c|}{n Nodes} & \multicolumn{2}{|c|}{LoRa School A (6 nodes)} & \multicolumn{2}{|c|}{LoRa School B (7 nodes)}\\
\hline
SF &  BW  &  Min. Period &  Max. Packets &   Min. Period &  Max. Packets&  Min. Period &  Max. Packets\\
   &  [kHz] &  [s] & [\#] &  [s] & [\#] &  [s] & [\#]\\
\hline
7 &  125 &  3.09*\textbf{\textit{n}} & 290.54/\textbf{\textit{n}} & 18.58 &  48.42 & 21.68 & 41.50\\
7 &  250 &  1.54*\textbf{\textit{n}}& 581.09/\textbf{\textit{n}} & 9.29 & 96.84 & 10.84 & 83.01\\
7 &  500 &  0.77*\textbf{\textit{n}} & 1162.19/\textbf{\textit{n}} & 4.64 & 193.69 & 5.42 & 166.02\\
9 &  125 &  12.39*\textbf{\textit{n}}& 72.63/\textbf{\textit{n}} & 74.34 &  12.10 & 86.73 & 10.37\\
9 &  250 &  6.19*\textbf{\textit{n}} & 145.27/\textbf{\textit{n}} & 37.17 & 24.21 & 43.36 & 20.75\\
9 &  500 &  3.09*\textbf{\textit{n}} & 290.54/\textbf{\textit{n}} & 18.58 & 48.42 & 21.68 & 41.50\\
\hline
\end{tabular}
\end{center}
\end{table}

The maximum \textit{packet rate} per node is achieved with \textit{SF} $7$ and \textit{BW} $500kHz$, which we implemented in our final network installation due to our priority of maximizing the sensing rate in the school and achieving a better sampling of the environmental reality in the public buildings. It is noteworthy that higher spreading factors allows for longer range at the expense of lower data rate, and vice versa.

We aim to compare the quality of the network under two extreme configurations. Configuration A provides higher rate (SF $7$, BW $500kHz$) and Configuration B provides longer range (SF $9$, BW $125kHz$). In order to study the network behaviour, we collected the following measurements per node: the number of \textit{DRPs} from the GW, the number of received \textit{DPs}, and the number of packets received with \textit{CRC} errors over a period of $15$ minutes. 

The maximum number of packets received under Configuration 1 is limited by the \textit{ToA} imposed by communications regulations. Thus we have to set the GW request rate accordingly to implement this restriction. As such, Configuration 1 is limited to $12$ packets, per $15$ minutes, per node. On the other hand, Configuration 2 is limited by the regulation at 193.69 (Table \ref{tab2}) packets, per $15$ minutes, per node. Effectively, Configuration 2 is restricted by the node design constraints. The GW requests data from each node after a $50 ms$ delay to guarantee the correct communication between the LoRa module and the micro-controller. In addition, the Environmental Nodes consume time to communicate through I2C with their digital sensors to collect the data for each request, limiting the final rate of the node. Due to these factors, in School Building A's installation every node can achieve a maximum of $174$ packets (Table \ref{tab3}) per $15$ minutes, which is lower but close to the theoretical limit. On both configurations, the average of the delivered \textit{DP} rate (Table \ref{tab3}) is higher for the nodes near the GW (Nodes 1, 5 and 6).

\begin{table}
\begin{center}
\caption{Number of delivered \textit{DPs} per node under different configurations in a 15 minute period}\label{tab3}
\begin{tabular}{|l|c|c|c|c|c|c|c|}
\hline
 & {} &   node1 &   node2 &   node3 &   node4 & node5 &   node6 \\
\hline
{\textbf{Configuration 1} }  & Avg.  &   11.17 &   11.17 &   11.15 &   11.16 &   11.16 &   11.16 \\
SF 9, BW 125 kHz & Min.   &    5.00 &    5.00 &    0.00 &    0.00 &    5.00 &    5.00 \\
& Max.   &   12.00 &   12.00 &   12.00 &   12.00 &   12.00 &   12.00 \\
\hline
{\textbf{Configuration 2} }  &  Avg.  &  155.24 &  149.77 &  145.79 &  150.87 &  155.23 &  155.22 \\
SF 7, BW 500 kHz & Min.   &   77.00 &    0.00 &    0.00 &    1.00 &   77.00 &   77.00 \\
& Max.   &  174.00 &  174.00 &  174.00 &  174.00 &  174.00 &  174.00 \\
\hline
\end{tabular}
\end{center}
\end{table}

As an indicator of the quality of the network, we define \textit{CRC Error Ratio} and \textit{Re-transmission Ratio}. \textit{CRC Error Ratio} is the ratio of \textit{DRPs} from the GW which resulted in a corrupted \textit{DP} being received. \textit{Re-transmission Ratio} is the ratio of \textit{DRPs} required to be repeated, either because of CRC errors, malformed \textit{DRP} or due to not receiving a reply. We are also interested in the connectivity between the GW and every node in our LoRa network. We quantify the quality of each link by calculating the \textit{Packet Delivery Ratio (PDR)} for every node. The PDR of the link between node A and GW can be measured as the ratio between the number of \textit{DPs} received by the GW from node A, and the number of \textit{DRPs} sent from the GW to the node A. The GW makes one \textit{DRP} and a maximum of 3 re-transmissions of the \textit{DRP} per node. In addition, we study the variation of the \textit{Received Signal Strength Indicator (RSSI)} per node in the network for both configurations.

\begin{table}
\begin{center}
\caption{Packet Delivery Ratio (PDR) per node under different configurations}\label{tab4}
\begin{tabular}{|l|c|c|c|c|c|c|c|}
\hline
 & {} &   node1 &   node2 &   node3 &   node4 & node5 &   node6 \\
\hline
\textbf{Configuration 1}  & Avg.  &    0.96 &    0.96 &    0.96 &    0.95 &    0.95 &    0.93 \\
SF 9, BW 125 kHz & SD   &    0.03 &    0.03 &    0.04 &    0.05 &    0.03 &    0.03 \\
& Min.   &    0.72 &    0.80 &    0.00 &    0.00 &    0.86 &    0.80 \\
& Max.   &    1.00 &    1.00 &    1.00 &    1.00 &    1.00 &    1.00 \\
\hline
\textbf{Configuration 2}   &  Avg. &    0.99 &    0.93 &    0.86 &    0.89 &    0.98 &    0.97 \\
SF 7, BW 500 kHz& SD   &    0.01 &    0.22 &    0.27 &    0.19 &    0.01 &    0.02 \\
& Min.   &    0.91 &    0.00 &    0.00 &    0.00 &    0.93 &    0.88 \\
& Max.   &    1.00 &    1.00 &    1.00 &    1.00 &    1.00 &    1.00 \\
\hline
\end{tabular}
\end{center}
\end{table}

\begin{table}
\begin{center}
\caption{Received Signal Strength Indicator (RSSI) per node under different configurations }\label{tab5}
\begin{tabular}{|l|c|c|c|c|c|c|c|}
\hline
& {} &   node1 &   node2 &   node3 &   node4 & node5 &   node6 \\
\hline
\textbf{Configuration 1}   & Avg. &  -45.06 &  -50.55 &  -87.78 &  -87.08 &  -53.52 &  -52.40 \\
Sf 9, BW 125 kHz & SD   &    0.41 &    0.98 &    2.59 &    1.73 &    1.35 &    1.29 \\
& Min.   &  -46.13 &  -54.25 &  -95.47 &  -95.29 &  -59.40 &  -55.77 \\
& Max.   &  -43.74 &  -48.75 &  -82.12 &  -84.12 &  -51.73 &  -50.17 \\
\hline
\textbf{Configuration 2}   &  Avg. &  -42.04 &  -46.29 &  -82.23 &  -86.86 &  -49.88 &  -44.78 \\
Sf 7, BW 500 kHz & SD   &    3.17 &   10.84 &   16.92 &    4.66 &    1.68 &    3.25 \\
& Min.   &  -55.33 &  -56.98 &  -89.00 &  -88.44 &  -57.70 &  -60.79 \\
& Max.   &  -37.40 &    0.00 &    0.00 &    0.00 &  -48.17 &  -41.27 \\
\hline
\end{tabular}
\end{center}
\end{table}

\begin{figure}
    {\centering
    \begin{subfigure}

        \includegraphics[width=0.49\textwidth]{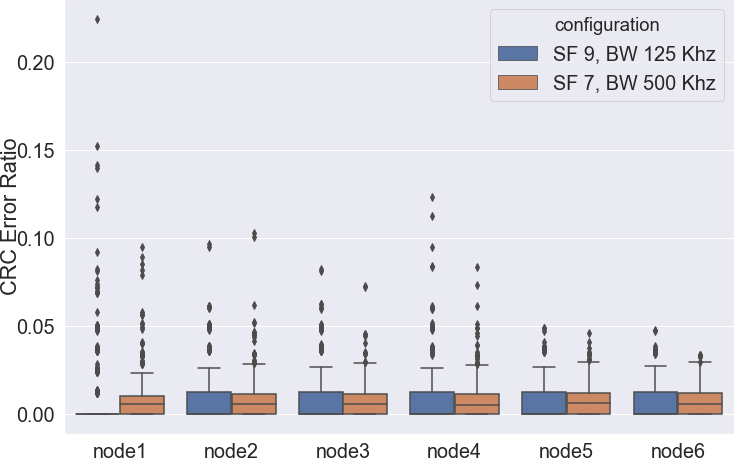}
        \label{fig:crc}
    \end{subfigure}
    \begin{subfigure}

        \includegraphics[width=0.49\textwidth]{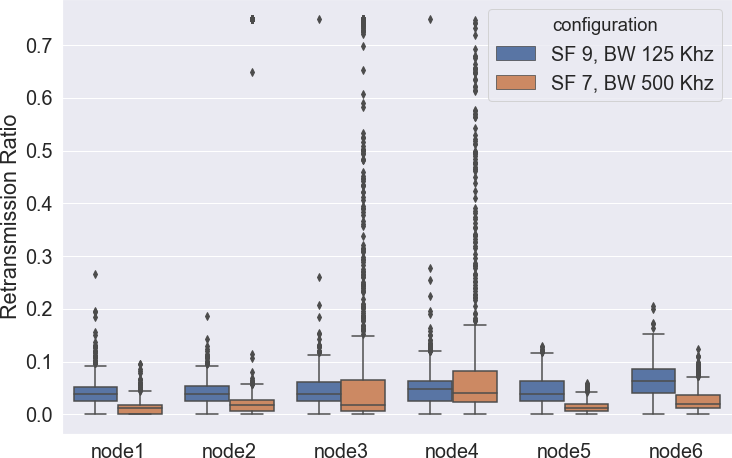}
        \label{fig:ret}
    \end{subfigure}}
    {\centering
    \begin{subfigure}

        \includegraphics[width=0.49\textwidth]{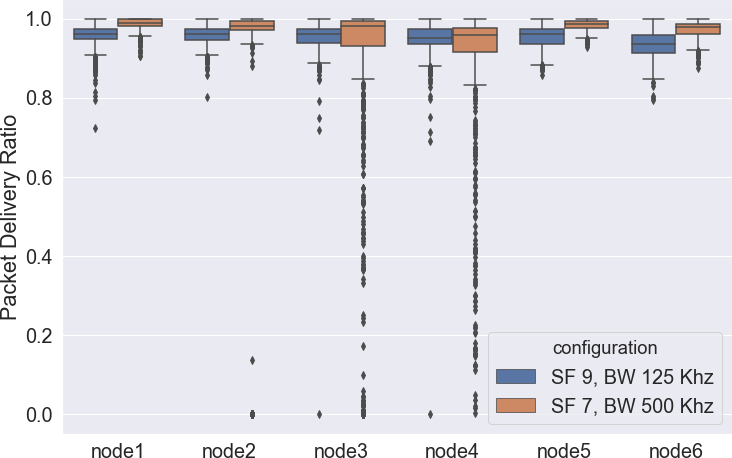}
        \label{fig:pdr}
    \end{subfigure}
    \begin{subfigure}

        \includegraphics[width=0.49\textwidth]{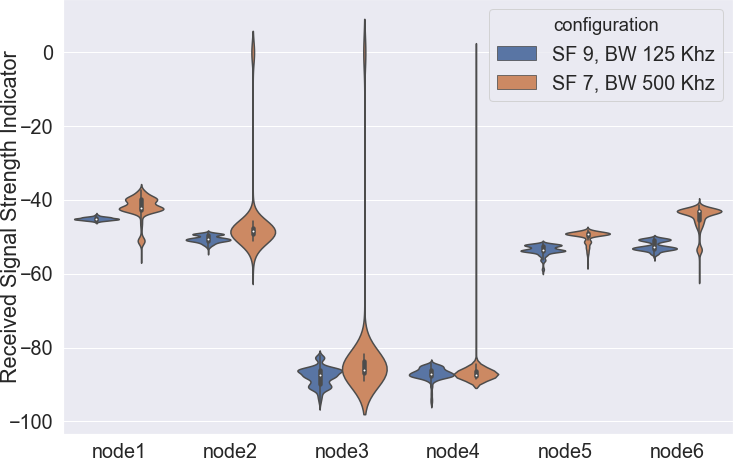}
        \label{fig:rssi}
    \end{subfigure}}
    \caption{Per node statistics for configurations \textit{\textbf{1:} SF 7, BW 500kHz} and \textit{\textbf{2:} SF 9, BW 125kHz}.}
    \label{fig:crc_ret_pdr_rssi}
\end{figure}

We expected to observe a worse quality network under Configuration 2, as a consequence of selecting parameter values that achieve a higher packet rate. We can observe that the median value for the \textit{CRC Error Ratio} distribution of each node is higher with similar standard deviation with the exception of the closest node (Node 1, CRC Error Ratio graph in Fig.~\ref{fig:crc_ret_pdr_rssi}). The degradation of network quality is evident at the farthest nodes (3 and 4) from the GW regarding the number of Re-transmissions (Re-transmission Ratio graph in Fig.~\ref{fig:crc_ret_pdr_rssi}), which exhibits higher standard deviation and more frequent and distant upper outliers. In addition, the PDR (Packet Delivery Ratio in Fig.~\ref{fig:crc_ret_pdr_rssi}) of these nodes is worse than in Configuration 1, with higher standard deviation and more frequent and distant lower outliers. On the other hand, the nodes closest to the GW achieve better network behavior regarding Packet Delivery and Re-transmission Ratios. Furthermore, the RSSI distribution (RSSI in Figure \ref{fig:crc_ret_pdr_rssi}) exhibits greater standard deviation, entailing less stable signal strength. 

In conclusion, we observed the expected cost in network quality, only in the further nodes, while in nearby nodes we observed an increase in the link's efficiency. This combined with the increase in the per node packet rate, resulted in a significant increase in sensor measurements across the whole network.

\section{XBee network behavior in school buildings}

Every node in the XBee network tries to deliver an \textit{Environmental Data Packet (EDP)} to the GW every $10$ seconds while emitting an extra \textit{PIR Data Packet (PDP)} each time motion is detected. The network overhead due to the extra \textit{PDPs} can saturate the medium and provoke a decrease of \textit{EDPs} delivered per node, thus decreasing the EDP rate. The data-set considered to analyze the specific behaviour in XBee School C is composed of the total number of packets delivered in the network in $5$ minute periods (\textit{EDPs} and \textit{PDPs} from every nodes). To quantify the effects of the independent \textit{PDPs}, we use their ratio against the observed maximum of the aggregation.

\begin{figure}
\includegraphics[width=\textwidth]{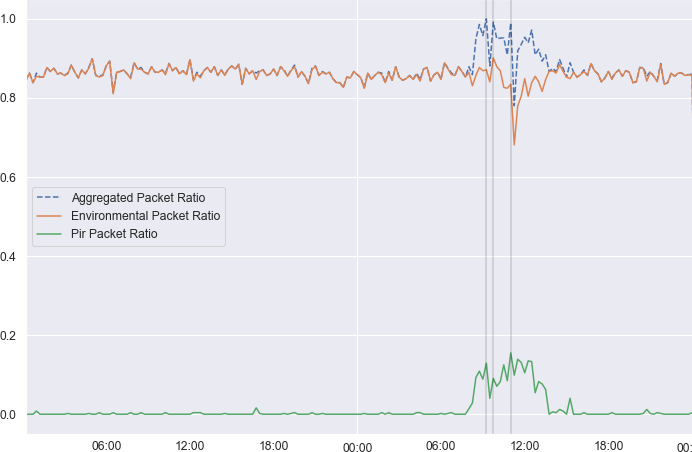}
\caption{\textit{Aggregated Packet Ratio}, \textit{EDP Ratio} and \textit{PDP Ratio} in XBee School C over 48 hours (Sunday and Monday)} \label{fig:pir}
\end{figure}

Fig.~\ref{fig:pir} shows clearly how during school hours the \textit{PDPs} can cause an observable decrease in the number of \textit{EDPs}, due to the saturation of the network at peak of \textit{PDP} Ratio. The maximum number of \textit{EDPs} is observed when the number of \textit{PDPs} is zero (Table \ref{tab6}). In addition, when the number of \textit{PDPs} exhibits a maximum, the number of \textit{EDPs} decreases below their average. The decision to include real-time motion detection to the network, can potentially be a hindrance to the stability of the \textit{EDP} rate of our network.

\begin{table}
\begin{center}
\caption{XBee Network behaviour. Number of \textit{Aggregated Packets}, \textit{PDPs} and \textit{EDPs} delivered at the time of maximum and minimum \textit{Aggregated Packet Ratio}, \textit{EDP Ratio} and \textit{PDP Ratio} respectively.}\label{tab6}
\begin{tabular}{|l||c||c|c||c|c||c|c|}
\hline
 \multirow{2}{*}{ } & \multirow{2}{*}{Average}    & \multicolumn{2}{|c||}{Aggregated Packets} & \multicolumn{2}{|c||}{PIR Packets} &  \multicolumn{2}{|c|}{Env. Packets} \\
 &  & Max & Min & Max & Min & Max & Min  \\
\hline
Aggregated Packets [\#]       & 143.05 & 185 & 84 &  178 &  142.11 &   157 &    87  \\
PIR Packets [\#]               &   1.91 &  41 &  3 &   45 &    0.00 &     0 &     7  \\
Node Packets [\#]              & 141.13 & 144 & 81 &  133 &  142.11 &   157 &    80  \\
\hline
\end{tabular}
\end{center}
\end{table}

\section{Discussion - Comparison inside school buildings}

\begin{figure}
    {\centering
    \begin{subfigure}

        \includegraphics[width=0.49\textwidth]{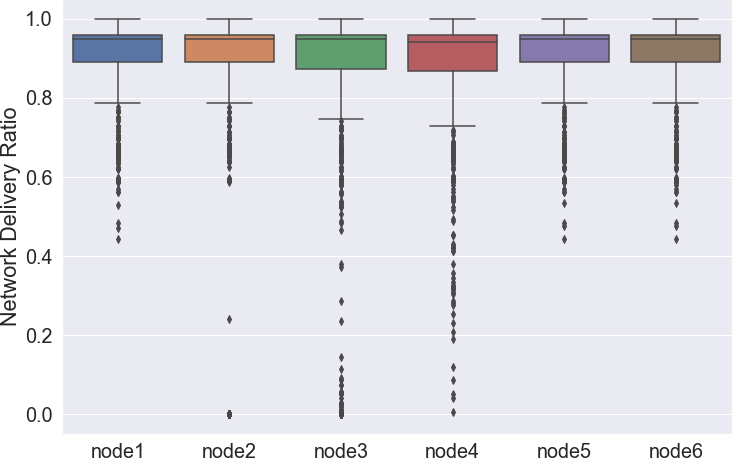}
        \label{fig:comp_lora_ndr}
    \end{subfigure}
    \begin{subfigure}

        \includegraphics[width=0.49\textwidth]{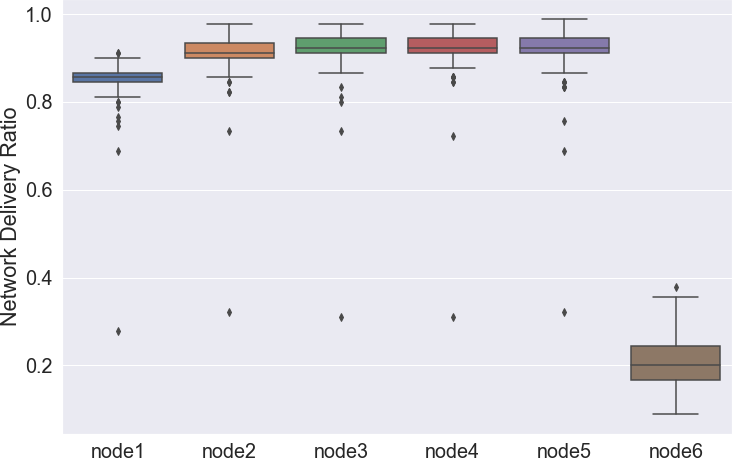}
        \label{fig:comp_xbee_ndr}
    \end{subfigure}}
    \caption{Network Delivery Ratio (NDR) per node in LoRa School A (left) under  Configuration 2 and XBee School C (right)}
    \label{fig:comp_lora_xbee}
\end{figure}

We aim to compare the quality of our LoRa and IEEE 802.15.4 IoT networks by comparing our observations from two real school buildings. LoRa School A consists of a LoRa Network with 6 nodes where the Node 3 and 4 are located at the farthest positions and the Node 1 and 5 at the nearest in relation with the GW. XBee School C has an XBee network consisting of 6 nodes, the farthest is node 6 and the nearest is the node 5. Due to the significant differences in radio and network architectures, to compare them we quantify the quality of these networks by the \textit{Network Delivery Ratio (NDR)}. The \textit{NDR} is defined as the ratio between the measured Delivered Packets and the potential maximum number of Packets that could be delivered by each node in the network in a time period, which in our case is 15 minutes. Every node in the XBee network is scheduled to attempt to send data to the GW every 10 seconds, resulting in a maximum of 90 packets in a 15 minute period. In the LoRa network using Configuration 2 and with 6 nodes, the network can achieve the delivery of a maximum of 174 packets. The data-set consists of \textit{NDR} measurements collected during a period of both business and weekend days for LoRa School A and XBee School C can be seen in Table \ref{tab7} and Figure \ref{fig:comp_lora_xbee}.     

We observe that the best network quality in terms of \textit{NDR} is observed in the LoRa School A under Configuration 1 which is the one with lowest rate where the maximum number of Data Packets per node in a 15 minute period is 12 packets. The network  in XBee School C exhibits a better \textit{NDR} than LoRa School A under Configuration 2 regarding the averages for every node with the exception of the farthest one that achieve a delivery of 20\% of the generated packets. On the other hand, the network in LoRa School A under Configuration 2 achieves a more stable \textit{NDR} across the installation, including the farthest nodes, with successful packet deliveries between 86\% and 89\% for every node and a significantly higher delivery rate.

The tree topology necessary for the XBee network to achieve comparable range to LoRa, influences negatively the packet rate of the nodes placed at the extremes of the tree. In comparison, LoRa's star network topology offered better coverage with a more stable data rate on all nodes.

\begin{table}
\begin{center}
\caption{Network Delivery Ratio (NDR) per node for LoRa and IEEE 802.15.4 Networks in different school buildings}\label{tab7}
\begin{tabular}{|l|c|c|c|c|c|c|c|}
\hline
Netwrok &{} &   node1 &   node2 &   node3 &   node4 & node5 &   node6 \\
\hline
\textbf{LoRa School A}  & Avg. &    0.94 &    0.94 &    0.94 &    0.94 &    0.94 &    0.94 \\
Conf. 1 & SD   &    0.03 &    0.03 &    0.04 &    0.04 &    0.03 &    0.03 \\
SF 9, BW 125 & Min.   &    0.41 &    0.41 &    0.00 &    0.00 &    0.41 &    0.41 \\
& Max.   &    1.00 &    1.00 &    1.00 &    1.00 &    1.00 &    1.00 \\
\hline
\textbf{LoRa School A}  &  Avg.  &    0.89 &    0.86 &    0.84 &    0.87 &    0.89 &    0.89 \\
Conf. 2& SD   &    0.12 &    0.22 &    0.25 &    0.17 &    0.12 &    0.12 \\
SF 7, BW 500 & Min.   &    0.44 &    0.00 &    0.00 &    0.01 &    0.44 &    0.44 \\
 & Max.   &    1.00 &    1.00 &    1.00 &    1.00 &    1.00 &    1.00 \\
\hline
\hline
\multirow{4}{*}{\textbf{XBee School C}}  &  Avg.  &    0.85 &    0.91 &    0.92 &    0.92 &    0.92 &    0.20 \\
& SD   &    0.04 &    0.05 &    0.05 &    0.04 &    0.05 &    0.05 \\
& Min.   &    0.28 &    0.32 &    0.31 &    0.31 &    0.32 &    0.09 \\
& Max.   &    0.91 &    0.98 &    0.98 &    0.98 &    0.99 &    0.38 \\
\hline
\end{tabular}
\end{center}
\end{table}

\section{Conclusions and Future Work}

Our work in recent years has resulted in the deployment of a large-scale IoT infrastructure inside a number of school buildings in Europe. In this context, we have opted to use different wireless networking technologies in order to test in practice their performance. With this work, we wanted to relay our practical experiences from using both IEEE 802.15.4 and LoRa for our specific application use-case and provide some practical examples and guidelines for IoT deployments that are similar to ours.

We have studied the behavior of both networks, in the scenario of changing the number of nodes in the network, varying the sampling rate of the sensors and the required data rate, or changing the distance between the IoT nodes inside the building. As an example of the results from the comparisons we made, LoRa decreases its delivery rate when increasing the number of nodes because of ToA European regulations which restricts the number of GW data requests in our network design. In 802.15.4 we expect an increased number of collisions when adding nodes due to CSMA. In the case of increasing the distance between nodes, LoRa achieves longer range with a stable rate, while 802.15.4 will need hop nodes in the middle, leading to increased number of collisions and an unstable rate in extreme nodes as a side effect.

Overall, our results show that in the use-case scenario and environmental settings of school buildings in Greece, LoRa-based wireless communication can have an advantage against competing technologies, in terms of reliability and complexity of networking. Regarding our future work, we plan to conduct a more thorough performance evaluation and explore in additional dimensions practical aspects like networking performance and reliability.

\bibliographystyle{splncs04}
\bibliography{paper}

\end{document}